\documentclass[a4paper,11pt]{article}
\pdfoutput=1 

\usepackage{jcappub} 

\usepackage[T1]{fontenc} 

\title{\boldmath Minimum length, extra dimensions, modified gravity and black hole remnants}

\author[]{Michael~Maziashvili}

\affiliation[]{Particle Physics \& Cosmology Group, Ilia State University,\\ 3/5 Cholokashvili Ave., Tbilisi 0162, Georgia}

\emailAdd{maziashvili@gmail.com}

\abstract{We construct a Hilbert space representation of minimum-length deformed uncertainty relation in presence of extra dimensions. Following this construction, we study corrections to the gravitational potential (back reaction on gravity) with the use of correspondingly modified propagator in presence of two (spatial) extra dimensions. Interestingly enough, for $r \rightarrow 0$ the gravitational force approaches zero and the horizon for modified Schwarzschild-Tangherlini space-time disappears when the mass approaches quantum-gravity energy scale. This result points out to the existence of zero-temperature black hole remnants in ADD brane-world model.}

\begin{document} 
\maketitle
\flushbottom

\section{Introduction}
\label{intro}

String theory has inspired a number of ideas which, have had a significant impact on particle physics model building and quantum gravity phenomenology \cite{Vafa:2009se, Polchinski:1998dz, Silagadze:1999gr}. One of the interesting predictions of string theory is that several extra dimensions must exist. Using the idea of large extra dimensions, Arkani-Hamed, Dimopoulos, and Dvali (ADD) suggested a phenomenological brane-world scenario for unifying the elementary particle forces with gravity near the electroweak energy scale \cite{ArkaniHamed:1998rs, ArkaniHamed:1998nn}. String theory, suggesting an unified description of gravity and elementary particles, becomes usually relevant only at very short distances - of the order of Planck length $l_P \simeq 10^{-33}$\,cm \cite{Witten:2002wb}, while ADD model opens the door for detecting some of the string theory ideas at distances comparable to the $\sim 10^{-17}$\,cm \cite{Antoniadis:1998ig}.

The other idea inspired by the string theory that received appreciable attention for studying the quantum gravity effects is the minimum-length deformed position-momentum uncertainty relation \cite{Veneziano:1986zf, Witten:1996} 

\begin{equation}\label{gur} \delta X \delta P  \,\geq \,
  \frac{\hbar}{2} \,+\, \beta G_N\delta P^2~,  \end{equation} where $\beta$ is a numerical factor of order unity and $G_N$ is the Newtonian coupling (from the very outset we assume $c=1$). A profound feature of Eq.\eqref{gur} is that it exhibits a lower bound on the position uncertainty of the order of $l_P=\sqrt{\hbar G_N}$. Apart from the string theory, the Eq.\eqref{gur} has a strong motivation from the black hole physics \cite{Maggiore:1993rv, Scardigli:1999jh}. Let us notice that the concept of minimum length in quantum gravity \cite{Mead:1964zz, Mead:1966zz, Garay:1994en} also hints at the modification of position-momentum uncertainty relation in such a way as to have the lower bound on position uncertainty.    

In what follows we will define both ingredients the brane-world and the minimum-length deformed position-momentum uncertainty relation in a purely phenomenological framework. On a dimensional grounds one might imagine lots of gravitational modifications to the uncertainty relation that manifest a nonzero lower bound on position uncertainty. (An early suggestion that gravity has the possibility of affecting the uncertainty relation was given by Mead \cite{Mead:1964zz, Mead:1966zz}). Though one of them is naturally singled out as it is discussed in the next section, we expect that the effect (back reaction) on gravity will have the same qualitative character for all minimum-length deformed quantum mechanics. The main purpose of this paper is indeed to provide a further example (apart from the one considered in \cite{Maziashvili:2011dx}) to see how a deformed quantum mechanics implying the existence of nonzero minimum position uncertainty affects the gravity.     

In ADD model \cite{ArkaniHamed:1998rs, ArkaniHamed:1998nn} the extra dimensions are compactified on $n$-torus. Matter fields are localized on the $3$ dimensional surface referred to as a (mem)brane and gravity is allowed to live in the whole space. The relation between the four and higher dimensional Newtonian couplings is 

\begin{equation}\label{relfourhighernewtonian} \text{Volume of extra space}\times G_N \,=\, \mathbb{G}_N~.  \nonumber  \end{equation} Thus, the quantum gravity energy scale

\[  \mathbb{E}_P \,=\,  \hbar^{(n+1)/(n+2)} \mathbb{G}_N^{-1/(n+2)} ~,  \] can be lowered at the expense of the volume of extra space. In such a scenario gravity becomes strong at the energy scale $\mathbb{E}_P$. Roughly, the gravitational potential on the ADD brane behaves as

 \begin{eqnarray}\label{spatialbox} 
   V(r)\, \simeq \,
   \begin{cases}
      -\mathbb{G}_N \, r^{-(1+n)} \, ,     &\text{if ~~ $ r \,\lesssim \, l_{ex} $~,}\\
      -G_N \, r^{-1} \,,          &\text{if ~~\, $r \,\gtrsim \, l_{ex}$~,}
   \end{cases} 
   \nonumber \end{eqnarray} where $l_{ex}$ denotes the size of extra dimensions. In what follows we will restrict ourselves to this semi-qualitative picture. The results obtained below are immediately applicable to the ADD model whenever the characteristic length scale of the process under consideration is $\ll l_{ex}$.

The ultimate goal of the present paper is to estimate the back-reaction on gravity due to minimum-length deformed quantum mechanics in presence of extra spatial dimensions and see how the Schwarzschild-Tangherlini space-time gets modified.

\section{Deformed quantum mechanics in $4+n$ dimensional space-time }

On dimensional grounds one can consider various gravitational corrections to the Heisenberg uncertainty relation that results in the lower bound on position uncertainty ($c=1$, that is, $[\hbar]=$g$\cdot$cm,\,$[\mathbb{G}_N]=$cm$^{n+1}/$g)

\begin{eqnarray}\label{gur1}\delta X \delta P \, &\geq & \, \frac{\hbar}{2} \,+\, \beta \, \hbar^{(\alpha -1)/\alpha} \mathbb{G}_N^{1/\alpha(n+1)}\delta P^{(n+2)/\alpha(n+1)} ~,~~ \nonumber \end{eqnarray} where $\beta$ is a numerical factor of order unity. In order to have a lower bound on position uncertainty, one should require

\begin{eqnarray}  \alpha \, \leq \, \frac{n+2}{n+1}~. \nonumber \end{eqnarray} 

On the other hand, it to be possible to switch off quantum mechanics $\hbar \rightarrow 0$, one has to require $\alpha \geq 1$. The choice $\alpha = 1$ is unique in that in this case the correction does not depend on $\hbar$ and therefore survives even when $\hbar \rightarrow 0$. By taking this specific choice one arrives at Eq.\eqref{gur} in absence of extra dimensions. So, the generalization of Eq.\eqref{gur} to the higher dimensional space takes the form

\begin{equation}\label{gurhigher} \delta X \delta P  \,\geq \,
  \frac{1}{2} \,+\, \beta \, \mathbb{G}_N^\frac{1}{n+1}\delta P^\frac{n+2}{n+1}~,  \end{equation}

\noindent (from now on we will adopt system of units $\hbar = c =1$).

The correction term in Eq.\eqref{gur} makes sense even when $\delta P \gtrsim G_N^{-1/2}$. In this case it is motivated by the fact that in high center of mass energy scattering, $\sqrt{s} \gtrsim G_N^{-1/2}$, the production of black holes dominates all perturbative processes \cite{Giddings:2001bu, Dvali:2010bf}, thus limiting the ability to probe short distances. (It is important to notice that at high energies, $\sqrt{s} \gg G_N^{-1/2}$, the black hole production is increasingly a long-distance, semi-classical process). To make the point clearer, the refined measurement of particle's position requires large energy transfer during a scattering process used for the measurement. But when the gravitational radius associated with this energy transfer $\sim G_N\sqrt{s}$ becomes grater than the impact parameter, the black hole will form and what one can say about the particle's position is that it was somewhere within the region $\sim G_N\sqrt{s}$. The gravitational radius of the black hole formed in the scattering process grows with energy as $r_g \simeq G_N \sqrt{s}$ determining therefore high energy behavior of the position uncertainty. 
  
Similar reasoning is valid in higher-dimensional case as well. The $4+n$ dimensional spherically symmetric gravitational field is described by the Schwarzschild-Tangherlini solution \cite{Tangherlini:1963bw, Myers:1986un}

\begin{eqnarray}  ds^2 \,=\, \left[ 1- \left(\frac{r_g}{r}\right)^{n+1} \right]dt^2 \,-\,   \left[ 1- \left(\frac{r_g}{r}\right)^{n+1} \right]^{-1}dr^2 \,-\, r^2 d\Omega^2_{n+2} ~,  \end{eqnarray} where $d\Omega^2_{n+2}$ is a line element of a $2+n$ dimensional unit sphere and the gravitational radius reads

\begin{eqnarray}\label{standardgradius}  r_g(m) \,=\, \left(\mathbb{G}_N m \right)^\frac{1}{n+1} \left[\frac{16 \pi }{(n+2)\text{Vol}\left(S^{n+2}\right)} \right]^\frac{1}{n+1} ~. \end{eqnarray} Thus we arrive at Eq.\eqref{gurhigher}; somewhat similar argument was used in \cite{Scardigli:2003kr} for deriving this equation.

\section{Hilbert space representation of uncertainty relation (\ref{gurhigher}) }
\label{Hilbert}

Here we closely follow the paper \cite{Kempf:1996nk}. In view of Eq.\eqref{gurhigher}, let us consider the deformed QM of the form \begin{equation}\label{hdmcomr} \left[ \widehat{X},\, \widehat{P}  \right]  \,=\, i \left(1 \, +\, \beta \, \mathbb{G}_N^\frac{1}{n+1} \widehat{P}\,^\frac{n+2}{n+1} \right) ~.  \end{equation} This sort of QM implies the existence of minimum position uncertainty of the order of \cite{Maslowski:2012aj}  

\[ \delta X \, \simeq \,  \left[\int\limits_0^{\infty} \frac{dP}{1 \, +\, \beta \, \mathbb{G}_N^\frac{1}{n+1} P\,^\frac{n+2}{n+1}} \right]^{-1} \,=\, \frac{\beta^\frac{n+1}{n+2}\mathbb{G}_N^\frac{1}{n+2}}{\int\limits_0^{\infty} \frac{dq}{1 \, +\, q\,^\frac{n+2}{n+1}} }~.  \]

A particular representation of $\widehat{\mathbf{X}},\, \widehat{\mathbf{P}}$ operators for a multidimensional generalization of Eq.\eqref{hdmcomr} 

\begin{eqnarray}\label{multdimcr}  
 \left[\widehat{X}_i,\, \widehat{P}_j \right] \,=\, i\left\{  \Xi\left(\widehat{P}^2\right) \delta_{ij} \,+\, \Theta \left(\widehat{P}^2\right)\widehat{P}_i\widehat{P}_j \right\}~,~ \left[\widehat{X}_i,\, \widehat{X}_j \right] \,=\, 0~,~ \left[\widehat{P}_i,\, \widehat{P}_j \right] \,=\, 0 ~,  
\end{eqnarray} can be constructed in terms of the standard $\widehat{\mathbf{x}},\, \widehat{\mathbf{p}}$ operators as\footnote{Latin indices take on the value $1, \ldots , 3+n$.  } 

\begin{eqnarray}\label{hspacerep} \widehat{X}_i \, = \, \widehat{x}_i \,=\, i\frac{\partial}{\partial p^i}~,  ~~ \widehat{P}_j \,=\,  \widehat{p}_j\xi\left(\widehat{\mathbf{p}}^2\right) \,=\, p_j\xi\left(\mathbf{p}^2\right)~. \end{eqnarray}
  
\noindent The simplest {\tt ansatz} would be to take

\begin{equation} \Theta \,=\, \frac{2\beta \,\mathbb{G}_N^\frac{1}{n+1}}{  \widehat{P}^\frac{n}{n+1}}  ~,\nonumber \end{equation} thus from Eqs.(\ref{multdimcr}, \ref{hspacerep}) we get  

\begin{eqnarray}&& \left( \frac{\partial}{\partial p^i} \, p_j\xi\left(\mathbf{p}^2\right) \,-\,  p_j\xi\left(\mathbf{p}^2\right) \frac{\partial}{\partial p^i}  \right)\psi(\mathbf{p}) \,=\,    \left( \delta_{ij}\xi\left(\mathbf{p}^2\right) \,+\,2p_ip_j \,\frac{d\xi\left(\mathbf{p}^2\right)}{d\mathbf{p}^2} \right)\psi(\mathbf{p})  \,=\, \nonumber \\&&  \left( \Xi\left(p^2\xi^2\right)\delta_{ij} +2\beta \, \mathbb{G}_N^\frac{1}{n+1} \,\frac{p_ip_j}{p^\frac{n}{n+1}} \,\xi^\frac{n+2}{n+1} \right)\psi(\mathbf{p})~, \nonumber  \end{eqnarray}

\noindent that is, 

\begin{eqnarray} \frac{d\xi\left(p^2\right)}{d p^2}  \,=\,   \beta \, \mathbb{G}_N^\frac{1}{n+1}  \,\frac{\xi^\frac{n+2}{n+1}}{p^\frac{n}{n+1}}~,~~ \Rightarrow ~~ \xi\left(p^2\right) \,=\, \left(1 \,-\,  \frac{2\beta}{n+2} \, \mathbb{G}_N^\frac{1}{n+1} p^\frac{n+2}{n+1}  \right)^{-(n+1)} ~. \nonumber \end{eqnarray}

\noindent Simplifying our notations by replacing

\begin{eqnarray}\label{gamartivebuliaghnishvna}  \frac{2\beta}{n+2} \, \mathbb{G}_N^\frac{1}{n+1}  \, \rightarrow \, \beta~,  \end{eqnarray} the above constructed representation takes the form 

\begin{eqnarray}\label{pdefqmrep}  \widehat{X}_i \, = \, \widehat{x}_i ~, ~~  \widehat{P}_j \,=\,  \widehat{p}_j\left(1 \,-\, \beta \, \widehat{p}^\frac{n+2}{n+1}  \right)^{-(n+1)} ~, \end{eqnarray} or in the eigen-representation of operator $\widehat{\mathbf{p}}$

\begin{eqnarray}\label{pdefqmrepeigenp}  \widehat{X}_j \, = \, i \, \frac{\partial}{\partial p_j} ~, ~~  \widehat{P}_j \,=\,  p_j\left(1 \,-\, \beta \, p\,^\frac{n+2}{n+1}  \right)^{-(n+1)} ~, \end{eqnarray} with the scalar product 

\begin{equation}\label{scalarproduct} \langle \psi_1 | \psi_2 \rangle = \int\limits_{p^{(2+n)/(n+1)} \,<\, \beta^{-1}}d^{3+n}p \, \,\psi^*_1(\mathbf{p})\psi_2(\mathbf{p})~. \nonumber \end{equation} In the case $n=0$ one recovers the result of \cite{Kempf:1996nk}.

Let us notice that the cutoff $p^{(2+n)/(n+1)} \,<\, \beta^{-1}$ arises merely from the fact that when $p$ runs over the region $p < \beta^{-(n+1)/(n+2)}$, $P$ covers the whole region from $0$ to $\infty$; see Eqs.(\ref{pdefqmrep}, \ref{pdefqmrepeigenp}). It might be instructive to look at this cutoff from the standpoint of Fourier transform. The standard uncertainty relation can be understood on the basis of Fourier transform since the spatial and momentum wave functions are related through it. The Fourier transform has the property that the more tightly localized the spatial wave function is, the less tightly localized the momentum function must be; and vice versa. Consequently, for in the minimum-length deformed quantum theory the momentum wave function can not spread beyond the region $\sim \beta^{-(n+1)/(n+2)}$, the spatial wave function can not be localized beneath the region $\sim \beta^{(n+1)/(n+2)}$.

\section{Imprints on field theory}

For simplicity let us consider a neutral scalar field. The modified field theory takes the form

\begin{eqnarray}\label{scaction}  \mathcal{A}[\varPhi] = - \int d^{4+n}x \, \frac{1}{2} \left[\varPhi\partial_t^2\varPhi  + \varPhi\widehat{\mathbf P}^2\varPhi 
+ m^2\varPhi^2 \right] ~, \end{eqnarray} that results in the equation of motion  

\begin{eqnarray}\label{eqofmot}
\left(\partial_t^2 \,+\, \widehat{\mathbf P}^2 \,+\, m^2 \right)\varPhi  \,=\,0~.
\end{eqnarray} Substituting a plane wave solution $\propto \exp\left(i[\mathbf{p}\mathbf{x} -\varepsilon_{\mathbf{p}}t]\right)$ in Eq.\eqref{eqofmot}, one finds the dispersion relation

\begin{eqnarray}\label{dispersionmodrel} \varepsilon^2 \,=\, \mathbf{P}^2 \,+\, m^2 \,=\, \frac{p^2}{\left(1-\beta p^\frac{n+2}{n+1}\right)^{2(n+1)}}  \,+\, m^2 \,=\,  \sum\limits_{n =0}^{\infty} (1+n)\beta^n p^{2(n+1)} +m^2~. \end{eqnarray}

\noindent In order to express $\mathbf{p}$ in terms of $\mathbf{P}$ one has to solve the equation 

\begin{eqnarray}  P^2 \,=\, \frac{p^2}{\left(1-\beta p^\frac{n+2}{n+1}\right)^{2(n+1)}}  ~, \nonumber \end{eqnarray} for $p$ and substitute it in  

\begin{eqnarray} \mathbf{p} \,=\,  \mathbf{P} \left[1-\beta p^\frac{n+2}{n+1}\left( P\right)\right]^{n+1}  ~. \nonumber  \end{eqnarray} The cut-off $p < \beta^{-(1+n)/(2+n)}$ readily indicates that the wave-length  

\begin{equation}\label{wavelength} \lambda \,=\, \frac{2\pi}{p\left(P\right)} ~, \nonumber \end{equation} is bounded from below $\lambda \geq 2\pi\beta^{(1+n)/(2+n)}$ no matter what the value of $P$ is.

The field operator takes the form

\begin{equation}\label{fieldoperator} \varPhi(t,\,\mathbf{x}) \,=\, \int\limits_{p^{(2+n)/(1+n)} < \beta^{-1}}\, \frac{d^{3+n}p}{\sqrt{ (2\pi)^{3+n}2 \varepsilon_\mathbf{p}}} \, \left[e^{i(\mathbf{p}\mathbf{x} - \varepsilon_\mathbf{p}t)}a(\mathbf{p}) \,+\, e^{-i(\mathbf{p}\mathbf{x} - \varepsilon_\mathbf{p}t)}a^+(\mathbf{p}) \right] ~,\end{equation} where 

\[\varepsilon_\mathbf{p} \,=\, \sqrt{ \frac{p^2}{\left(1-\beta p^{\frac{2+n}{1+n}} \right)^{2(n+1)}}  \,+ \,  m^2 }~.\]

From Eqs.(\ref{scaction}, \ref{fieldoperator}) one gets

\begin{eqnarray}\label{Hamiltonian}&& \mathcal{H} =   \frac{1}{2} \int d^{3+n}x \, \left[\varPi^2  + \varPhi\widehat{\mathbf P}^2\varPhi 
+ m^2\varPhi^2 \right] \,=\, \nonumber \\&&  \int\limits_{p^{(2+n)/(1+n)} < \beta^{-1}}d^{3+n}p \, \frac{  \varepsilon_\mathbf{p}}{2} \left[a^+(\mathbf{p})a(\mathbf{p}) \,+\, a(\mathbf{p})a^+(\mathbf{p})\right] ~. \nonumber  \end{eqnarray}

The field quantization condition 

\begin{eqnarray}\label{fieldmodiiedcommutator} \left[\varPi(\mathbf{x}), \, \varPhi(\mathbf{y})\right] \,=\, -i \int\limits_{p^{(2+n)/(1+n)} < \beta^{-1}}d^{3+n}p \,\, \frac{e^{i\mathbf{p}(\mathbf{x} - \mathbf{y})}}{ (2\pi)^{3+n}} ~, \nonumber  \end{eqnarray} is now imposed by the smeared-out delta function. Roughly, this means that $\varPi(\mathbf{x})$ and $\varPhi(\mathbf{y})$ operators do not commute for $|\mathbf{x} -\mathbf{y}| \lesssim \beta^{(1+n)/(2+n)}$. Let us notice that quantum field theories with a minimal length scale have also been studied in \cite{Hossenfelder:2003jz}, and they were shown in \cite{Hossenfelder:2007fy} to yield a modification of the equal time commutation relation very similar to the one found here.

\section{Back reaction on gravity and black hole remnants }

The corrected potential (calculated by the modified propagator with respect to Eq.\eqref{dispersionmodrel})

\begin{eqnarray}\label{corrpotmuldim} V(r) \,=\,  \frac{\text{Vol}\left(S^{n+2}\right)}{(2\pi)^{3+n}} \int\limits_{k^{(2+n)/(n+1)} \,<\, \beta^{-1}} d^{3+n}k \, \left(\frac{1}{k^{1/(n+1)}} \,-\, \beta \,k \right)^{2(n+1)} e^{i\mathbf{k}\mathbf{r}} \,=\, \nonumber \\ \frac{\text{Vol}\left(S^{n+2}\right)}{(2\pi)^{3+n}}  \int\limits_{k^{(2+n)/(n+1)} \,<\, \beta^{-1}} d^{3+n}k \, \left[ \frac{1}{k^2} \,-\,   \frac{2(n+1)\beta }{k^{n/(n+1)}} \,+\, (n+1)(2n+1)\beta^2 k^{2/(n+1)}   \,+\,  \right. \nonumber \\  \left. \ldots \,-\,2(n+1) \beta^{2n+1}k^{(2n^2+3n)/(n+1)} \,+\, \beta^{2(n+1)}k^{2(n+1)} \right] e^{i\mathbf{k}\mathbf{r}} ~,~~ \end{eqnarray} results in the modified Schwarzschild-Tangherlini space-time

\begin{eqnarray}  ds^2 \,=\, \left[ 1 \,-\, r_g^{n+1} V(r) \right]dt^2 \,-\,    \left[ 1 \,-\, r_g^{n+1}V(r) \right]^{-1}dr^2 \,-\, r^2 d\Omega^2_{n+2} ~, \end{eqnarray} where $r_g$ is given by Eq.\eqref{standardgradius}.

From now on we will consider a specific case $n=2$. In this case \begin{equation}\label{gravradiusioridamganz}
r_g \,=\, \left(\frac{3 \mathbb{G}_N m}{2\pi}\right)^{1/3} ~.
\end{equation} For $n=2$, the Eq.\eqref{corrpotmuldim} takes the form (see \ref{damatebameore} Appendix)

\begin{eqnarray}V(r) \,=\, \frac{2}{3\pi\beta^{9/4}}  \int\limits_0^{1}  d\tilde{k}   \left(\tilde{k}^2 \,-\,  6\tilde{k}^{10/3} \,+\, 15\tilde{k}^{14/3}   \,-\, 20\tilde{k}^6 \,+\, 15\tilde{k}^{22/3} \,-\,   6\tilde{k}^{26/3}   \,+\, \tilde{k}^{10}   \right) \times \nonumber \\ \left[ \frac{\sin(\tilde{k}\tilde{r})}{(\tilde{k}\tilde{r})^3} \,-\, \frac{\cos(\tilde{k}\tilde{r})}{(\tilde{k}\tilde{r})^2} \right] ~,\nonumber   \end{eqnarray} $\left(\tilde{k}=k\beta^{3/4},\, \tilde{r} =r\beta^{-3/4}\right)$ with the asymptotic behavior (see Eq.\eqref{mtsiremandzilebze} in the \ref{damatebameore} Appendix)

\begin{eqnarray}\label{potentsialisasimptyofaktseva} V\left(r \ll \beta^{3/4}\right) \,=\,  \frac{2}{3\pi \beta^{9/4}}   \left[ 0.00295112  \,+\, 0.0000393787 \, \frac{r^2}{\beta^{3/2}}  \,+\, 3.0709\times 10^{-7} \, \frac{r^4}{\beta^3} \,+\, \right. \nonumber \\ \left.  1.65633\times 10^{-9} \, \frac{r^6}{\beta^{9/2}}  \,+\, \cdots \right]  ~.  ~~\end{eqnarray}

\begin{figure}[h]

\includegraphics[width=0.8\textwidth,origin=c]{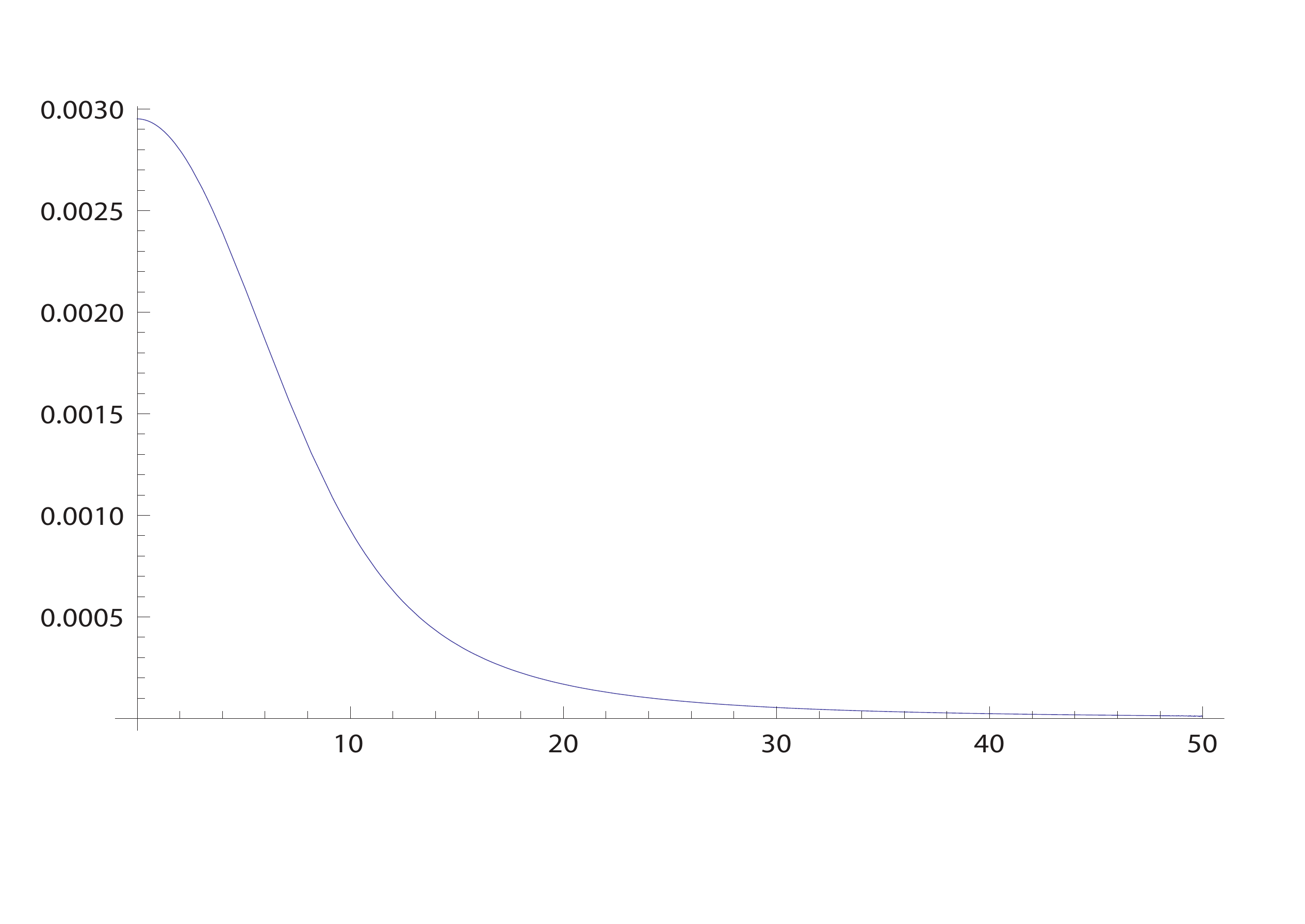}\\
\caption{ The graphic of the function $1.5\pi \widetilde{V}(r)$; on the horizontal axis $r$ is measured in units of $\beta^{3/4}$. }
\end{figure}

\noindent  The equation for the gravitational radius now reads 

\begin{eqnarray}\label{modschtangradiusi}  \frac{\beta^{9/4}}{r_g^3} \,=\, \widetilde{V}(r)~, \end{eqnarray} where $\widetilde{V}(r) = \beta^{9/4}V(r)$. Observing that $\widetilde{V}(r)$ is a monotonically decreasing function, see Fig.1, with its maximum $\widetilde{V}(0) = 0.00196741/\pi$, one infers that the solution of Eq.\eqref{modschtangradiusi} approaches zero when the black hole mass approaches  

\begin{equation}\label{shavikhvrelisnarchenismasa} m_{remnant} \,=\, \frac{\pi^2\beta^{9/4}}{0.00295112 \times \mathbb{G}_N} \,\rightarrow\, \frac{\pi^2\beta^{9/4}\mathbb{G}_N^{-1/4}}{2^{9/4}\times 0.00295112} ~, \end{equation} where we have used Eqs.(\ref{gravradiusioridamganz}, \ref{gamartivebuliaghnishvna}). That is, in the last expression of Eq.\eqref{shavikhvrelisnarchenismasa} $\beta$ is a dimensionless parameter of order unity. Thus, when black hole evaporates to the mass, $m_{remnant}$, the horizon disappears. For masses smaller than $m_{remnant}$, the Eq.\eqref{modschtangradiusi} does not have any solution.

Recalling that Hawking temperature is proportional to the surface gravity, one infers that black hole emission temperature approaches zero when black hole evaporates down to the $m_{remnant}$ (we use Eq.\eqref{potentsialisasimptyofaktseva}) 

\begin{equation}\label{shavikhvrelinarchenistemperetaura} T_{H}(m_{remnant}) \,\propto \, \left.\frac{dV(r)}{dr}\right|_{r=0} \,=\, 0~. \end{equation}

\section{Discussion}

We have constructed the Hilbert space representation for a higher-dimensional minimum-length deformed position-momentum uncertainty relation, which was originally suggested in \cite{Scardigli:2003kr}. In absence of extra dimensions one recovers well known example studied in \cite{Kempf:1996nk}.

The minimum-length deformed quantum mechanics implies corrections for the both sides of the Einstein equation

\begin{equation}\label{Einstein}  R_{\mu\nu} -\frac{g_{\mu\nu}R}{2} = 8\pi G_N T_{\mu\nu} ~.\end{equation} The energy-momentum tensor of the matter fields is now understood to be modified with respect to the minimum-length deformed field theory (modifications arise both at first- and second-quantization levels, for details see \cite{Berger:2010pj, Maziashvili:2011qs}). It is hard to imagine what might be the analogue of the minimum-length deformed field theory for the gravitational field, to figure out the corresponding corrections to the left-hand side of Eq.\eqref{Einstein}. Nevertheless, one might proceed with a semi-classical description; the disturbance (graviton field) can be separated off from the background space-time for which the minimum-length deformed field formalism can be applied immediately. Let us notice that this kind of approach is most straightforward for formulating a quantum theory of gravity \cite{Faddeev:1973zb} (in this paper, the expansion of the gravitational field around the flat space-time was used that allows one to view gravity as QFT for self-interacting spin-2 field). It is also worth noticing that despite the non-renormalizability, the method of expansion of the gravitational field around the fixed background enables to make meaningful predictions for graviton radiative corrections in the framework of an effective field theory approach \cite{Khriplovich:2002bt, Khriplovich:2004cx, Kirilin:2006en, BjerrumBohr:2002kt, BjerrumBohr:2002ks, Burgess:2003jk}.

Two sources of the corrections to the Newtonian potential can be identified. The first concerns the existence of maximally localized states in the framework of minimum-length deformed quantum mechanics that naturally replaces the delta-function distribution for a point-like particle and alters the Poisson equation (see \cite{Maziashvili:2011dx}). Correspondingly, one can take some sort of smeared-out delta function instead of delta-function distribution, which describes the point-like source and find its gravitational field to get some idea about the minimum-length modified black holes \cite{Nicolini:2005vd, Ansoldi:2008jw, Spallucci:2011rn}.

 Another source altering the Newtonian potential is the modified dispersion relation that enters the propagator and thus provides corrections to the right-hand side of Eq.\eqref{Einstein} (in the linearized theory). That is, the ($3+n$ dimensional) Laplace operator entering the standard equation of motion gets modified: $\widehat{\mathbf{p}}^2 \rightarrow \widehat{\mathbf{P}}^2$ and correspondingly the Poisson equation for a point like particle $\widehat{\mathbf{p}}^2 V(\mathbf{r}) \sim \delta(\mathbf{r})$ is altered as: $\widehat{\mathbf{P}}^2 V(\mathbf{r}) \sim \delta(\mathbf{r})$. In addition one has to take into account that the latter equation should be solved in the class of functions admitting a cutoff Fourier representation. Following this way, as in the four-dimensional case \cite{Maziashvili:2011dx}, the modified gravity shows up the following interesting features, gravitational force vanishes when $r\rightarrow 0$ (see Eq.\eqref{potentsialisasimptyofaktseva}); the horizon disappears when the black hole evaporates down to the Planck mass (see Eqs.(\ref{modschtangradiusi}, \ref{shavikhvrelisnarchenismasa})) and the corresponding Hawking temperature vanishes (see Eq.\eqref{shavikhvrelinarchenistemperetaura}). Here let us mention the following papers \cite{Tkachuk:2007zz, Helling:2007zv, AmelinoCamelia:2010rm} where the corrections to the potential due to deformed dispersion relations are estimated but without cutoff on $p$. The cutoff is important for it has an implicit reference to modified uncertainty relation implying the minimum position uncertainty.

The question of black hole remnants in the framework of generalized uncertainty relation was originally addressed in a heuristic way in \cite{Adler:2001vs}. The basic idea put forward in this paper is to identify the black hole radius with the position uncertainty of emitted particle and estimate its characteristic momentum/energy scale by using the minimum-length deformed position-momentum uncertainty relation. Black hole radiation temperature defined this way results in the zero heat capacity when the black hole evaporates down to the Planck mass that might be understood as the reason for the radiation shut off (the temperature itself becomes complex quantity when the black
hole mass drops below the Planck scale). Following this paper, the extra-dimensional/braneworld corrections to the black hole radiation due to Eq.\eqref{gur} (and also corrections to the black hole production cross section) have been studied in many papers; see some of them: \cite{Cavaglia:2003qk, AmelinoCamelia:2005ik, Bolen:2004sq, Cavaglia:2004jw, Hossenfelder:2004ze, Koch:2005ks, Nouicer:2007pu, Xiang:2009yq, Stetsko:2012kx, Nozari:2012ng}.

Let us briefly comment on the validity of approximation we have used. As it is well known, in the framework of GR the linearized approximation is bad for describing both the near horizon structure (whatever the size of the black hole is) and the region of singularity. In both cases one might get an idea that gravity becomes enormously strong just by using the Newtonian potential (which comes from the free propagator); in the former case it is manifested by the fact that the escape velocity becomes equal to $c$, and in the latter case it becomes merely infinitely strong. (In field theory terminology - near the horizon, nonlinearities of the equation of motion manifest themselves; while approaching the Planck length, gravity enters the strong regime and quantum gravitational effects become essential). But in the present case the key point is that gravity becomes weak in the near-horizon region for the Planck scale black holes; the surface gravity vanishes when black hole mass approaches the Planck scale (see Eq.\eqref{shavikhvrelinarchenistemperetaura}). This argument is certainly just an attempt to get an intuition and does not provide any rigorous motivation. As long as we do not know how to implement the minimum-length deformed quantum mechanics in GR (beyond the linearized case) it is impossible to be rigorous about this point.

For the same reason we can say nothing about the quantum (radiative) corrections to the propagator for we do not know what the full implementation of minimum-length deformed quantum mechanics in GR might look like.

Finally let us note that the above constructed representation for minimum-length deformed quantum mechanics in presence of extra dimensions can be used for estimating corrections to various quantities and processes in braneworld scenario \cite{Hossenfelder:2004wv, Hossenfelder:2004gj, Bhattacharyya:2004dy, Bhattacharyya:2005hq, Mimasu:2011sa, Mimasu:2012xy}.

\appendix

\section{Appendix}
\label{damatebameore}

To estimate the integral

\begin{eqnarray}\label{appint}
\int\limits_{k^{(2+n)/(n+1)} \,<\, \beta^{-1}} d^{3+n}k \, \left[ \frac{1}{k^2} \,-\,   \frac{2(n+1)\beta }{k^{n/(n+1)}} \,+\, (n+1)(2n+1)\beta^2 k^{2/(n+1)}   \,+\, \ldots \right. \nonumber \\ \left. \,+\, \beta^{2(n+1)}k^{2(n+1)} \right] e^{i\mathbf{k}\mathbf{r}}  ~, \end{eqnarray}

\noindent let us choose the axis $x_{3+n}$ along $\mathbf{k}$ and introduce spherical coordinates in the momentum space 

\[ k_1 \,=\, k\sin\varphi \prod\limits_{j=1}^{n+1} \sin\theta_j ~,~~ k_2 \,=\, k\cos\varphi \prod\limits_{j=1}^{n+1} \sin\theta_j ~,~~k_{i+2} \,=\, k\cos\theta_i \prod\limits_{j=i}^{n+1} \sin\theta_j~,~~ k_{3+n}  \,=\, k\cos\theta_{n+1}~,\] 

\noindent where $i = 1, \ldots , n~;~k \geq 0~,~ 0\leq \varphi < 2\pi~,~ 0 \leq \theta_j \leq \pi$. Thus, we get $\mathbf{k}\mathbf{r} = k x_{3+n} \cos\theta_{n+1}~,~d^{3+n}k = k^{2+n}\,dk\,d \varphi \prod\limits_{j=1}^{n+1} \sin^j\theta_j \, d\theta_j $, and the integral \eqref{appint} reduces to   

\begin{eqnarray}&&\text{Vol}\left(S^{n+1}\right)  \int\limits_0^{\beta^{-(1+n)/(2+n)}}  dk   \left(k^n \,-\,   2(n+1)\beta k^{(n^2 + 2n +2)/(n+1)} \,+\, (n+1)(2n+1)\beta^2 k^{(n^2 +3n +4)/(n+1)}   \,+\, \right.\nonumber \\&& \left.  \ldots  \,-\,   2(n+1) \beta^{2n+1}k^{(3n^2+6n+2)/(n+1)}   \,+\, \beta^{2(n+1)}k^{3n+4}   \right)\int\limits_0^{\pi} d\theta_{n+1} \, \sin^{n+1}\theta_{n+1}  \, e^{ik\, x_{3+n}\cos\theta_{n+1}} ~.\nonumber \end{eqnarray}

\noindent In the specific case $n=2$ we get the following expression. Denoting $\cos\theta_3 = t$ one finds 

\begin{eqnarray}  \int\limits_0^{\pi} d\theta_{3} \, \sin^3\theta_{3}  \, e^{ikx_{5}\cos\theta_{3}} \,=\, \int\limits_{-1}^1dt (1-t^2)e^{ik\,x_5t} ~. \nonumber \end{eqnarray} Knowing the integral

\begin{eqnarray} I(k\, x_{5})  \,=\, \int\limits_{-1}^1dt\, e^{ik\,x_5t} \,=\, \frac{2\sin(k\,x_5)}{k\,x_5} ~,  \nonumber  \end{eqnarray} one readily estimates  

\begin{eqnarray} \frac{d^2I(k\, x_{5}) }{d(k\, x_{5})^2} \,=\,-\, \int\limits_{-1}^1dt\,t^2 e^{ik\,x_5t} ~,  \nonumber  \end{eqnarray} that results in

\begin{eqnarray}   \int\limits_{-1}^1dt (1-t^2)e^{ik\,x_5t} \,=\, \frac{4\sin(k\,x_5)}{(k\,x_5)^3} \,-\, \frac{4\cos(k\,x_5)}{(k\,x_5)^2} ~. \nonumber \end{eqnarray} So, the modified potential takes the form

\begin{eqnarray}&&V(r) \,=\,  \frac{\text{Vol}\left(S^4\right) \text{Vol}\left(S^3\right) }{(2\pi)^5}  \int\limits_0^{\beta^{-3/4}}  dk   \left(k^2 \,-\,  6\beta k^{10/3} \,+\, 15\beta^2 k^{14/3}   \,-\, 20\beta^3k^6 \,+\, 15\beta^4 k^{22/3} \,-\, \right.  \nonumber \\&& \left.   6 \beta^5k^{26/3}   \,+\, \beta^6 k^{10}   \right) \,   \left[ \frac{4\sin(kr)}{(kr)^3} \,-\, \frac{4\cos(kr)}{(kr)^2} \right] \,=\,   \frac{4 \text{Vol}\left(S^4\right) \text{Vol}\left(S^3\right) }{(2\pi)^5 \beta^{9/4}}  \int\limits_0^{1}  d\tilde{k}   \left(\tilde{k}^2 \,-\,  6\tilde{k}^{10/3}  \right.  \nonumber \\&& \left.  \,+\, 15\tilde{k}^{14/3}   \,-\, 20\tilde{k}^6 \,+\, 15\tilde{k}^{22/3} \,-\,   6\tilde{k}^{26/3}   \,+\, \tilde{k}^{10}   \right) \,     \left[ \frac{\sin(\tilde{k}\tilde{r})}{(\tilde{k}\tilde{r})^3} \,-\, \frac{\cos(\tilde{k}\tilde{r})}{(\tilde{k}\tilde{r})^2} \right] ~,\nonumber   \end{eqnarray} where the dimensionless quantities $\tilde{k},\, \tilde{r}$ are defined as $k = \tilde{k}\beta^{-3/4},\, r= \tilde{r}\beta^{3/4}$. The behavior of this potential for distances $r \ll \beta^{3/4}$ can be obtained by using the asymptotic expression 

\[ \frac{\sin(x)}{x^3} \,-\, \frac{\cos(x)}{x^2} \,=\, \frac{1}{3} \,-\,\frac{x^2}{5\cdot 3!} \,+\, \frac{x^4}{7\cdot 5!} \,-\, \frac{x^6}{9\cdot 7!} \,+\, \ldots ~, \] that gives

\begin{eqnarray}\label{mtsiremandzilebze}&& V\left(r \ll \beta^{3/4}\right) \,=\,  \frac{4 \text{Vol}\left(S^4\right) \text{Vol}\left(S^3\right) }{(2\pi)^5 \beta^{9/4}}   \left[ 0.00295112  \,+\, 0.0000393787 \, \frac{r^2}{\beta^{3/2}}  \,+\, \right. \nonumber \\&&  \left. 3.0709\times 10^{-7} \frac{r^4}{\beta^3} \,+\,  1.65633\times 10^{-9} \frac{r^6}{\beta^{9/2}}  \,+\, \cdots \right]  ~. ~~ \end{eqnarray}

\acknowledgments

Stimulating comments from Micheal~S.~Berger and Zurab~K.~Silagadze are kindly acknowledged. The work was supported in part by the grant: FR/498/6-200/12.


\end{document}